\documentclass[aps,prl,twocolumn,groupedaddress]{revtex4}
\usepackage{amsfonts,amssymb}
\usepackage{amsthm}
\usepackage{tikz}
\usetikzlibrary{calc}
\usetikzlibrary{arrows}
\usepackage{tikz-3dplot}
\usepackage{color} 
\usepackage[fleqn]{amsmath}
\usepackage{amsthm}
\usepackage{amssymb}
\usepackage{amsfonts}
\usepackage{bbm}
\usepackage{epsfig}
\usepackage{times}
\usepackage{hyperref}
\usepackage{bm}
\usepackage{float} 
\usepackage{appendix} 
\usepackage{mathrsfs}
\usepackage{appendix}

\newcommand*{\braopket}[3]{\langle #1| #2 | #3\rangle}

\usepackage{graphicx}

\DeclareGraphicsExtensions{.png}

\newcommand{\comment}[1]{}
\newcommand{\ket}[1]{\left |  #1 \right\rangle}
\newcommand{\bra}[1]{\left \langle #1  \right |}

\theoremstyle{plain}
\newtheorem{theorem}{Theorem}

\theoremstyle{definition}

\begin{document}

	\title{Knowledge-Concealing Evidencing of Knowledge about a Quantum State}

\author{Emily \surname{Adlam}} \affiliation{Centre for Quantum
	Information and Foundations, DAMTP, Centre for Mathematical
	Sciences, University of Cambridge, Wilberforce Road, Cambridge, CB3
	0WA, U.K.}  \author{Adrian \surname{Kent}} \affiliation{Centre for
	Quantum Information and Foundations, DAMTP, Centre for Mathematical
	Sciences, University of Cambridge, Wilberforce Road, Cambridge, CB3
	0WA, U.K.}  \affiliation{Perimeter Institute for Theoretical
	Physics, 31 Caroline Street North, Waterloo, ON N2L 2Y5, Canada.}
	\date{\today}

\begin{abstract} Bob has a black box that emits a single pure
state qudit which is, from his perspective, 
uniformly distributed.   Alice 
wishes to give Bob evidence that she has knowledge about the emitted state
while giving him little or no information about it.
We show that zero-knowledge evidencing of such
knowledge is impossible in quantum relativistic protocols, 
extending a previous result of Horodecki et
al..  We also show that no such protocol can be both sound and complete. 
We present a new quantum relativistic protocol which we conjecture to be close
to optimal in security against Alice and which reveals little
knowledge to Bob, for large dimension $d$.  We
analyse its security against general attacks by Bob
and restricted attacks by Alice.   
\end{abstract}
	
		\maketitle
	
	\section{Introduction}
	
Zero-knowledge proving is a cryptographic primitive in which
one agent proves a fact to another agent without giving away
any information other than that the fact is true. It has a
wide range of practical applications, particularly in
electronic voting schemes \cite{evoting} and digital signature
schemes \cite{dsignature}, and is also used for a variety of
theoretical purposes, such as showing that a language is easy
to prove \cite{Fortnow}. Zero-knowledge proving of
\emph{knowledge}, where Alice is required to prove only that she
knows some fact, without giving Bob any information about the
fact itself, is a particularly useful version of this task
which plays a key role in a number of identification protocols
\cite{Feige}.
		
Horodecki et al.\@ \cite{HorodeckiZKP}
explored the possibility of what they called a ``zero knowledge convincing protocol
on quantum bit''. 
In their model, a verifier (henceforth called Bob) knows he has 
a single copy of a pure 
qubit, but has no other information about the state.   
A prover (henceforth called Alice)
wishes to make a prediction that Bob can verify and
that will hold with certainty
only if she knows what the state is,  
without giving Bob any additional information about its identity. 
They showed that no non-relativistic protocol involving classical
information exchanges and quantum Alice-to-Bob communications can 
implement this task securely \cite{HorodeckiZKP}.
They also discussed  
some protocols that implement very weak versions of the
task, either giving Bob a great deal of information about
the qubit, or giving him only weak evidence of Alice's knowledge,
or both.   

Horodecki et al.\@'s pioneering discussion was 
informal on some points. 
It did not fully distinguish cases in which Alice has classical
knowledge about Bob's quantum state (e.g. a classical data string
describing it) from cases in which she has quantum knowledge (e.g. a box able to make only some
fixed number of copies).     Nor did it underline that Alice cannot
{\it prove} that she knows a precise 
classical description of a single quantum state even if she is not
concerned about 
about giving Bob information.  This is because the classical information
about the state that can be extracted by measurement is bounded,
and Alice always has a boundedly nonzero chance of guessing this
information even if she knows nothing about the state.   (See Theorem
2 below.)  Alice may also have a high chance of guessing the information even
if she has only partial information about the quantum state -- for example, she can predict the outcome of a complete projective measurement
on a qubit with probability $\frac{1}{2}$. 
Similarly, even if she only knows a dimension $2$ subspace in which a qudit
lies, she can still specify a complete projective measurement whose outcome
she can predict with probability $\frac{1}{2}$. 
Moreover, if the state is $\eta$ and Alice believes it is $\eta'$,
where the fidelity $F(\eta, \eta')$ is close to $1$, then she is 
almost as likely to pass any protocol testing her knowledge of $\eta$ 
as she would be if she knew $\eta$.   
Since Bob only has a single copy of the state, Alice cannot provide
more evidence by repeating a protocol that tests her knowledge. 
By contrast, in classical contexts Alice's chances of success can typically be made 
arbitrarily small by iteration, so that one can
reasonably (modulo epsilonics) speak of classical zero knowledge
\emph{proofs}.  

Other interesting questions left open include: What can relativistic 
protocols achieve?  How much evidence can Alice provide?   
What bounds exist on the tradeoffs 
between the evidence Alice provides and the amount of
knowledge she gives away?  Are there protocols strong enough for
practical cryptographic purposes?   How do the answers depend
on the dimension $d$ of the state space?   

We explore these questions below, beginning 
with some formal definitions.  
We prove a stronger no-go theorem showing that 
no protocol that provides non-trivial evidence of Alice's 
knowledge about a pure quantum state of finite
dimension can prevent Bob from acquiring some additional
knowledge about the state, even in the setting of relativistic
quantum cryptography. 
We also prove a bound on the strength of evidence Alice can provide.   
Since proofs of knowledge of a finite-dimensional quantum state are
not possible, and zero-knowledge protocols that give some evidence
of knowledge are also not possible, we then consider the weaker
but feasible task of \emph{knowledge-concealing evidencing of knowledge about a quantum state}
(KCEKQS).  

A KCEKQS protocol requires 
Alice to give Bob evidence that she has some 
form of knowledge about a quantum state whilst giving him 
incomplete information about the state.  
Ideally, a successfully completed protocol should
give Bob as much evidence as possible, without assuming Alice's honesty.
Ideally, too, the protocol should ensure as small a bound as possible on the
information obtainable by Bob, whether or not it is successfully
completed or he honestly follows it.    We discuss some simple protocols, generalising 
protocols previously considered by Horodecki et
al. \cite{HorodeckiZKP}, and show that they are relatively weak 
in knowledge-concealment, or in evidencing knowledge, or both.  
We then propose a new relativistic quantum protocol. 
We show it is secure against restricted attacks by Alice and
general attacks by Bob, for large $d$, in a sense we make precise below.
We conjecture this remains true for general attacks by
Alice.
	
\section{Definitions}

We assume for now that Alice has no option to 
abort the protocol; allowing an abort option does not significantly 
change our main results \cite{sm}. 
	
A non-relativistic \emph{knowledge-concealing evidencing of knowledge about a quantum state
(KCEKQS)} protocol involves two mistrustful parties, Alice and Bob,
occupying disjoint secure laboratories.  We assume that each party has trusted error-free
devices in their own laboratory; both parties trust error-free classical and
quantum communication channels between the laboratories; we consider 
errors and losses later \cite{sm}.  
Bob begins in possession of a quantum system $Q_B$
known to be prepared in some pure state $\eta = \ket{\eta} \bra{\eta}$ drawn uniformly 
at random from $Q_B$.   The protocol 
requires Alice and Bob to act on alternate rounds and terminates
after a fixed finite number of rounds. 
Each round may require a party to carry out unitary operations
and/or measurements on a quantum system 
in their possession and/or to send classical and/or quantum
information to the other party.   The protocol may specify that these
actions are probabilistically determined, according to given
probability distributions.     
The final round of the protocol requires Bob to generate one of two possible outcomes,
$0$ and $1$, from the classical and quantum information in his
possession.   These correspond to Bob's rejecting or accepting that 
Alice has provided evidence of knowledge about $\eta$.    
We write $p(0)$ and $p(1)$ for the outcome probabilities.  

In a relativistic KCEKQS protocol, each party may have several trusted
agents occupying separate secure laboratories, with secure
communications between them, lying within pre-agreed regions. 
One agent of Bob's initially possesses $Q_B$.   The protocol
requires Alice's and Bob's agents to carry out unitary operations
and/or measurements on quantum systems in their possession and to send classical and/or quantum
communications to given other agents of the same party and/or the other
party, within their agreed location regions and within agreed time intervals.   
The protocol may specify these
actions are probabilistically determined, according to given
probability distributions.     
The protocol terminates after a fixed finite number of such actions.
The final prescribed action is for one of Bob's agents to 
generate one of two possible outcomes,
$0$ and $1$, from the classical and quantum information in his
possession, as above. 

We will characterise the efficiency of an ideal KCEKQS protocol by three parameters
$\epsilon_{\rm C}$, $\epsilon_{\rm K}$ and $\epsilon_{\rm S}$; we 
discuss other relevant features of KCEKQS protocols in the
supplementary material \cite{sm}.   
When evaluating these parameters for specific protocols, we will
mostly consider the ideal case of error-free devices and channels.  
In realistic implementations, channel noise, device
errors and losses may alter the parameter values.\footnote{Typically, 
uncorrected noise, errors and losses will increase $\epsilon_{\rm C}$ and decrease
$\epsilon_{\rm K}$, and uncorrected losses will decrease $\epsilon_{\rm S}$.}   
However, our no-go theorems below still hold in reasonable models of
noise, errors and losses, so long as these are uncorrelated with
$\eta$ and with any knowledge Alice may have of or about $\eta$.  

We define our parameters by the following criteria, in each case
averaging
over $\eta$: 
\begin{itemize}
			
\item \textbf{Completeness:} If Alice has a precise classical
description of $\eta$ and both parties perform the protocol correctly, then $p(1) = 
1 - \epsilon_{\rm C}$.   
			
\item \textbf{Soundness:} If Alice has no classical or quantum
information about the state $\eta$, then $\epsilon_{\rm S}$ is 
the supremum of $p(1)$ over all possible (honest or dishonest)
strategies for Alice, assuming that Bob performs the protocol correctly.  
			
\item \textbf{Knowledge-concealing:} Suppose that Alice performs the protocol
correctly and at the start of the protocol Bob knows
nothing about the state $\eta$.    Then $\epsilon_{\rm K}$ is the supremum
of the expected squared fidelity $F^2 ( \eta, \phi ) = | \, \langle \eta | \phi
\rangle \, |^2$ over (honest or dishonest) strategies that give Bob the value
of a pure state $\phi$ as a guess for $\eta$.  

\end{itemize} 			

Let $\epsilon_{\rm M}$ be the
supremum of the same expected squared fidelity obtainable by Bob if he does not 
take part in the protocol and carries out quantum operations and
measurements on $Q_B$.  
We call $\epsilon_{\rm K} - \epsilon_{\rm M}$ the \emph{knowledge
  gain} available from the protocol to a dishonest Bob.  
We say the protocol is \emph{ zero-knowledge}
if $\epsilon_{\rm K} = \epsilon_{\rm M}$,   We say it is \emph{non-trivial} if $1 - \epsilon_{\rm C} > \epsilon_{\rm S}$.   
			
As defined above, general protocols allow both classical and quantum
communications in both directions.   We will also consider 
examples with more restricted communications. 
Extending the discussion of Ref. \cite{HorodeckiZKP}, we consider
\emph{classical} protocols, in which Alice and Bob employ only
classical communication, \emph{quantum A-to-B}
protocols, which additionally allow quantum communications from A to
B, and similarly \emph{quantum B-to-A} protocols. 
	
\section{No-go theorems}

Horodecki et al.\@$~$\cite{HorodeckiZKP} showed that no non-relativistic 
KCEKQS classical or quantum A-to-B protocol for an unknown qubit 
has $\epsilon_{\rm C} =0$, $\epsilon_{\rm
  S}<1$, and is zero-knowledge.    
We establish here a considerably more general result, applying to 
relativistic KCEKQS protocols for qudits with two-way classical
and/or quantum communications, 
and general parameter values.   Our statements apply
to protocols whose security is based only on quantum theory and
special relativity, i.e. within the standard scenario for
unconditionally secure relativistic quantum cryptography \cite{kentrelfinite}. 

\begin{theorem} 

There exists no non-trivial zero-knowledge KCEKQS protocol. \cite{sm}

\end{theorem} 

We also establish a tradeoff between completeness and soundness which 
bounds the degree of evidence Alice can provide:
 
\begin{theorem} 
For any qudit KCEKQS protocol, $\frac{\epsilon_{\rm S}}{1 -
  \epsilon_{\rm C}} \geq \frac{1}{d}$.  \cite{sm}
\label{zeroknowledgetheorem}
\end{theorem}

In particular, theorem \ref{zeroknowledgetheorem} means that for small $d$, $\epsilon_{\rm S}$ and $\epsilon_{\rm C}$ cannot both be close
to $0$, regardless of the value of $\epsilon_{\rm K}$.
This makes the case of large $d$ particularly interesting to explore.

We observe that the bound of theorem \ref{zeroknowledgetheorem}
is tight. 
For 
example, it is attained by a protocol in which Alice predicts to Bob
the outcome of a projective measurement that includes $\eta$ on
the system $Q_B$: this has $\epsilon_{\rm S} = \frac{1}{d}$ and
$\epsilon_{\rm C}=0$.  More generally, it is attained for a protocol
in which Alice is required to predict this outcome and also predict the
outcome of some independent random event with success probability $p$:
this has $\epsilon_{\rm S} = \frac{p}{d}$ and $\epsilon_{\rm C}=
1-p$. 
We say a KCEKQS qudit protocol is {\it CS-optimal} if $\epsilon_{\rm S} = \frac{1}{d},
\epsilon_{\rm C} = 0$. 

\section{Protocols }
	
\subsection{Classical A-to-B \label{classicalmessage}}

As noted above, Horodecki et al.\@ \cite{HorodeckiZKP} argue that
non-trivial non-relativistic zero-knowledge classical A-to-B 
protocols with $\epsilon_{\rm C} = 0$ are impossible for a qubit.
In any such protocol, Alice must predict some measurement outcome,
and any measurement prediction that holds with certainty for 
a pure qubit $\eta$ and is not certain for a random qubit 
allows Bob to identify $\eta$ exactly,
and so has $\epsilon_{\rm K} = 1$. 

One might instead consider protocols with
$\epsilon_{\rm C} > 0$, in which Alice chooses a projective 
measurement which includes a randomly chosen projector $P$ from those
with $\braopket{\eta}{P}{\eta} = 1 - \epsilon_{\rm C}$.
One might also consider strengthening such protocols
by allowing Alice to use a secure relativistic bit commitment
\cite{kentrel,kentrelfinite, bcsummoning, AdlamKent1, AdlamKent2}
to commit her predicted outcome, unveiling this commitment if and
only if Bob's reported outcome agrees with her prediction.
However, such protocols still have either $\epsilon_{\rm K}$ or 
$\epsilon_{\rm C}$ large \cite{sm}. 
			
\subsection{Quantum A-to-B \label{quantummessage}}

Horodecki et al.\@ \cite{HorodeckiZKP} also consider a protocol where
Alice gives Bob a copy of $\eta$; as they note, such protocols
can achieve 
$\epsilon_{\rm C} = 0$ and 
$\epsilon_{\rm K} < 1$.
Indeed it is possible to achieve $\epsilon_{\rm K} \ll 1$ for large $d$.  In this
sense, for large $d$, their protocol outperforms the classical A-to-B protocols
just discussed. 
However, one needs to consider the tradeoffs
between $\epsilon_{\rm K}, \epsilon_{\rm C}$ and $\epsilon_{\rm S}$.  
It is also worth highlighting that this protocol requires
Alice only to possess quantum information about $\eta$ rather
than classical information.   Alice can ensure $p(1)=1$  
even if she only has a black box that will make only a single copy 
of $\eta$ and has no other classical or quantum information about
$\eta$.    
	
We extend the discussion of Ref. \cite{HorodeckiZKP}
by considering a generalisation of their protocol in which
Alice gives Bob $N$ copes of $\eta$: 

\begin{enumerate} 
	
	\item 	Alice prepares $N$ systems $\{ S_i \}$ in the state $\eta$ and gives them to Bob. 
	
	\item Bob performs a measurement $\{ \Pi_S, \mathbb{I} - \Pi_S\}$
where $\Pi_S$ is the projector onto the symmetric
subspace of the joint state space of the $\{ S_i \}$ and
$Q_B$.\footnote{A motivation for this choice is that, given`a system in
state $\psi^{\otimes n}$ and another system in state
$\phi^{\otimes m}$, for some integers $m, n$, the measurement
$\{ \Pi_S, \mathbb{I} - \Pi_S\}$ (i) always gives outcome $1$ if
$\psi = \phi$ (ii) maximises the probability of outcome $0$ if $\psi
\neq \phi$, among measurements satisfying (i) \cite{Barnett, ensemblecomparison}.}
	
	\item If the result is $\Pi_S$, Bob accepts; otherwise he
          rejects. 
 
\end{enumerate} 

If Alice knows $\eta$ and follows the protocol, Bob will accept, so
this protocol achieves $\epsilon_{\rm C} = 0$. 
We show \cite{sm} that
 $\epsilon_{\rm S}=\frac{1}{N + 1} + \frac{N}{d(N + 1)}$.
So, for $N=1$, we have $\epsilon_{\rm S} = \frac{1}{2} +
\frac{1}{2d}$, while, for $N$ large,  $\epsilon_{\rm S} \rightarrow \frac{1}{d}$, 
so the protocol tends to CS-optimality in this limit.
However, we also show \cite{sm} that $\epsilon_{\rm K} = 
\frac{N + 2}{N + d+1}$, which tends to $1$ for large $N$,
while $\epsilon_{\rm M} = \frac{2}{d+1}$.   
Thus $\epsilon_{\rm K} >  \frac{1}{d
\epsilon_{\rm S}}$  and $\epsilon_{\rm K} - \epsilon_{\rm M} >
\frac{d-1}{d+1} \frac{N}{N+2} \frac{1}{d \epsilon_{\rm S}}$ 
for all $d,N$, which are relatively poor tradeoffs. 
In particular, near 
CS-optimality ($ \epsilon_{\rm S} \approx \frac{1}{d}$) 
implies near-zero concealment ($\epsilon_{\rm K} \approx 1$)
and implies significant knowledge gain ($\epsilon_{\rm K} - \epsilon_{\rm M}
\approx \frac{d-1}{d+1}$).   
 
\subsection{A quantum B-to-A protocol \label{quantmvp}} 

We now propose a new relativistic protocol that involves quantum B-to-A
and two way classical communication.   
Given certain security assumptions, 
we show that,
the protocol asymptotically tends to CS-optimality in the 
limit where the security parameter $N$ is large, taking
the parameter $q = \lceil \frac{N}{d} \rceil $.
We conjecture that this remains true without the relevant assumptions.

We show that $ \epsilon_{\rm K} \leq \frac{4}{d+1}$, while 
$\epsilon_{\rm M} = \frac{2}{d+1}$, so 
$\epsilon_{\rm K} \rightarrow 0$ and 
$\epsilon_{\rm K} - \epsilon_{\rm M} \rightarrow 0$ 
for large $d$ \cite{sm}. 

\begin{enumerate} 
	
\item Alice and Bob agree in advance on positive integer
security parameters $N$ and $q$. 
	
\item Bob prepares $N $ quantum systems $\{ S_i \}$ in states chosen
uniformly at random.
	
\item Bob randomly permutes the systems $\{ S_i \}$ and the system
$Q_B$, assigns them all indices from $1$ to $N + 1$, and then gives
all $N + 1$ systems, labelled by their indices, to Alice.
	
\item Alice carries out the projective measurement
    $\{ \eta  \, , \,  \mathbb{I} -  \eta \}$ on each of the $N + 1$ systems that
  Bob gave her.   Write $C'$ for the list of indices for
which she obtains outcome $ \eta$; 
let $| C' | = q'$.  If $q' \leq q$, she forms a list $C = C' \cup D$, 
where $D$ is a list of $(q-q')$ copies of the dummy index $0$.\footnote{This dummy index
prevents cheating strategies in which Bob uses the number of Alice's
commitments, made at the next step, to extract additional 
information about the state.}.  If $q' > q$, she picks a random size $q$
sublist $C$ of $C'$.   
	
\item Alice randomly permutes $C$ and then
performs $q$ relativistic bit string commitments \cite{sm} committing her
to each of the indices in the permuted list.  Each bit string commitment is set up
so that Alice can commit to any index in $\{0, 1, 2,  \ldots N+1\}$.   
	
\item Bob tells Alice the index $x \in \{ 1, \ldots, N+1 \}$ that he assigned to $Q_B$.
	
\item If $x \in C$, Alice unveils her commitment to that index.  
Otherwise she announces failure and Bob rejects.\footnote{In the
ideal error-free case, if Alice knows $\eta$ precisely and both
parties honestly perform the protocol, failure is 
possible if and only if $q'>q$.} 
	
\item If Alice's unveiled commitment is indeed $x$, Bob accepts.   
Otherwise he rejects. 
	
\end{enumerate} 

We analyse the security of this protocol in the supplementary 
material.   We show the bound on $\epsilon_{\rm K}$ above holds for
any attacks by Bob, and that $\epsilon_{\rm C} \rightarrow 0$ 
for $q = \lfloor \frac{N}{d} \rfloor$ in the large $N$ limit. 
We also show that $\epsilon_{\rm S} \approx \frac{1}{d}$ if Alice is 
restricted to strategies in which she commits to $q$ classical
values chosen from $\{ 0 , \ldots , N+1 \}$ and unveils one of
these committed values.   
A full security analysis requires a complete analysis of general
quantum operations Alice could carry out to produce unveiling
data; we leave this for future work. 

\section{Conclusion} 
	
We have proven two no-go theorems demonstrating that even in the
relativistic setting there is no perfect KCEKQS protocol for quantum
states of finite dimension and bounding the evidence Alice can
supply. 
We have also described a 
new protocol involving quantum Bob-to-Alice communications
and relativistic signalling constraints, which appears to achieve a
significant improvement on existing protocols for large $d$. 
Although it is not zero knowledge for finite $d$, 
it reveals little extra information to Bob for large $d$.    
We conjecture that it is asymptotically CS-optimal, i.e. offers 
essentially optimal security against Alice.   
We anticipate that this protocol may be a valuable quantum 
cryptographic primitive in contexts where marginal revelations 
of information to Bob are acceptable.    	

{\bf Acknowledgments} \qquad This work was partially supported by an FQXi grant and by
Perimeter Institute for Theoretical Physics. Research at Perimeter
Institute is supported by the Government of Canada through Industry
Canada and by the Province of Ontario through the Ministry of Research
and Innovation.

\end{document}


\title{Supplementary Information for
Knowledge-Concealing Evidencing of Knowledge about a Quantum State}
		
		\author{Emily \surname{Adlam}} \affiliation{Centre for Quantum
			Information and Foundations, DAMTP, Centre for Mathematical
			Sciences, University of Cambridge, Wilberforce Road, Cambridge, CB3
			0WA, U.K.}  \author{Adrian \surname{Kent}} \affiliation{Centre for
			Quantum Information and Foundations, DAMTP, Centre for Mathematical
			Sciences, University of Cambridge, Wilberforce Road, Cambridge, CB3
			0WA, U.K.}  \affiliation{Perimeter Institute for Theoretical
			Physics, 31 Caroline Street North, Waterloo, ON N2L 2Y5, Canada.}
		\date{\today}
		
		\maketitle

\section{Protocols with an abort option} 

Before proving our main results, we want to extend our definitions
to allow for the possibility that Alice may abort the
protocol.  In a relativistic KCEKQS protocol with abort option, each party may have several trusted
agents occupying separate secure laboratories, with secure
communications between them, lying within pre-agreed regions. 
One agent of Bob's initially possesses $Q_{\rm B}$.   The protocol
requires Alice's and Bob's agents to carry out unitary operations
and/or measurements on quantum systems in their possession and to send classical and/or quantum
communications to given other agents of the same party and/or the other
party, within their agreed location regions and within agreed time
intervals.   The protocol may specify these
actions are probabilistically determined, according to given
probability distributions.\footnote{One could also allow protocols
that specify nondeterministic actions but do not specify probability
distributions, as considered for bit commitment by D'Ariano et
al.\cite{d2007reexamination}.  We do not consider such protocols here.} 

After each round of receiving data, each of Alice's agents generates
one of two possible outcomes, $0$ and $1$, from the classical and quantum information in her
possession.   If she gets outcome $0$, she announces to Bob's neighbouring agents, within a
pre-agreed time interval, that the protocol
is aborted.   Bob's agent communicates this to Bob's other agents.  
Alice's agent also announces the abort to all of Alice's other agents.   Any agent who
receives an abort message stops participating in the protocol from that
point.  

If there is no abort, the protocol terminates after a fixed finite
number of communications between Alice's and Bob's agents.    
The allowed timings of abort announcements and of Bob's final
announcement are fixed so that, in 
the event of any valid abort announcement, no agent of Bob's will
announce the protocol outcome.  
If there is no abort, then the final prescribed action is for one of Bob's agents to 
generate one of two possible outcomes,
$0$ and $1$, from the classical and quantum information in his
possession.  These correspond to Bob rejecting or accepting that 
Alice has provided evidence of knowledge of $\eta$.    
The possible outcomes are thus disjoint events $0,1,{\rm abort}$. 
We write $p(0),p(1),p({\rm abort})$ for the outcome probabilities.  

A non-relativistic protocol with abort option proceeds similarly, with
Alice having the option to announce an abort after each Bob-to-Alice
communication before Bob's final announcement.    

\vspace{5mm}
{\bf Non-triviality for protocols with an abort option:} \qquad 

We define $\epsilon_{\rm C}$, $\epsilon_{\rm K}$ and $\epsilon_{\rm
  S}$ as before.    
We say a protocol with abort is
\emph{non-trivial} if $1 - \epsilon_{\rm C} > \epsilon_{\rm S}$, as
before. 

\textbf{Comments:} 1. Our definition of non-triviality is intended to 
characterise what might reasonably be considered to be a useful
knowledge-evidencing protocol.    
Consider two possible hypotheses.  The first is that  
Alice has no classical or quantum information correlated with $\eta$. 
The second is that Alice knows the classical value of $\eta$ precisely
and follows the protocol honestly.  
If the protocol allows Alice to attain a higher or equal value of
$p(1)$ when the first hypothesis holds than the value attained
when the second holds, then outcome $1$ gives Bob no evidence to prefer the 
second hypothesis over the first.   Our non-triviality condition excludes this
possibility, meaning that outcome $1$ gives Bob at least some
evidence to prefer the second hypothesis over the first.
(Of course, there are other possible hypotheses.  For example, Alice could 
have some definite classical information about $\eta$, or some
beliefs about $\eta$ that she expresses in a probability distribution,
or some quantum information correlated with $\eta$, or combinations
of these.   Non-triviality does not necessarily imply that the outcome
$1$ gives Bob evidence in favour of the hypothesis that Alice knows
$\eta$ precisely and follows the protocol honestly compared to any
of these hypotheses.)
  
Outcome $1$ is the only outcome that can sensibly have this property.
Bob can never rely on an abort as 
giving evidence that Alice knows $\eta$, since a dishonest
Alice can always abort.  And it makes no sense for an
outcome $0$ to give positive evidence that Alice knows $\eta$ compared
to outcome $1$: if this were the case then the protocol outcomes should be
relabelled.        

2. In fact, a weaker definition of non-triviality suffices for
our no-go theorem below.   Let $p(x | \psi ; \eta)$ be the probability
distribution for the three outcomes $x \in \{0,1,{\rm abort} \}$
when the state of $Q_{\rm B}$ is $\psi$, Bob performs the protocol correctly, and Alice performs
the version of the protocol which would be correct if she knew that
the state of $Q_{\rm B}$ were $\eta$.  Then it is sufficient that 
$p( x | \psi ; \eta)$ depends non-trivially on $\psi$ for fixed
$\eta$. 

\section{No-Go Theorems }

\subsection{Zero-knowledge} 

\begin{theorem} 

There exists no non-trivial KCEKQS protocol which is zero-knowledge. 

\end{theorem} 

\begin{proof} 

Consider a non-trivial KCEKQS protocol $\mathscr{P}$ applied to 
a system $Q_{\rm B}$ with Hilbert space $\mathscr{H}_B$.
In any such protocol Bob may replace the unknown state $\eta$ 
by any state of his choice.  Suppose that he does so, but that the protocol is otherwise
honestly performed by both parties, with Alice acting in the
belief that the unknown state is $\eta$. 
The protocol then defines a quantum measurement operation $P^{\eta}$ on 
$Q_{\rm B}$.   Now if Bob inputs a randomly chosen pure state $\phi \in Q_{\rm B}$, 
non-triviality and continuity imply that
there is a neighbourhood $N$ of $\phi$ with nonzero measure $\delta$ 
(with respect to the uniform measure), such that 
$$
P ( \eta \in N \, | \, {\rm outcome~}1 \, ) > \delta \, . 
$$
Nontriviality and continuity also imply that $\epsilon_{\rm C} < 1$ and that 
$P ( \, {\rm outcome~}1 \, ) > 0$.  
Hence, with nonzero probability, Bob gains some
nontrivial information about the distribution of $\eta$ from
this operation.
Since he retains $\eta$, he also can carry out any measurement
he wishes on $\eta$.   Combining information from the two
processes gives him on average strictly more information than is available
from the measurement alone.   Hence $\epsilon_{\rm K} > \epsilon_{\rm M}$ 
and the protocol is not zero-knowledge.    
	
\end{proof} 

\subsection{Completeness vs soundness}

\begin{theorem} 
In a KCEKQS protocol for a pure quantum state in a Hilbert space of
dimension $d$, the completeness and soundness parameters obey 
$$\frac{\epsilon_{\rm S}}{1 - \epsilon_{\rm
      C}} \geq \frac{1}{d} \, . 
$$
	
\label{zeroknowledgetheorem}
\end{theorem}
\begin{proof} 

A protocol $P$ for KCEKQS may require either party to carry
out measurements, to make random choices from a classical
probability distribution, to introduce quantum states, and/or to send classical data.
For any such protocol, we can define a related fully
quantum protocol $QP$
in which all data are introduced as quantum ancillae at the start
and kept at the quantum level until the outcomes
are obtained.   Thus, to define $QP$ from $P$, measurements are replaced
by unitary measurement interactions (without extracting
measurement data), classical random choices are replaced
by interactions with entangled ``quantum dice'' (without
extracting data about the dice outcome) and classical
communications are replaced by quantum communications 
of states in a pre-agreed orthonormal basis (for instance the 
computational basis).   We take the protocol $QP$ to
proceed thus until the last steps.  The first of these is for all
the information in Alice's possession to be sent to one of Alice's
agents, who carries out a measurement giving her
the outcome ${\rm abort}$ or ${\rm not~abort}$. 
She communicates this outcome to all of Alice's and all of Bob's
agents, within an agreed time window, so that Bob's agents all know
the outcome after a prespecified coordinate time. 
If no abort is communicated to any of Bob's agents within the
prescribed time, they send all the information in their possession
to one of Bob's agents, who carries out a single
measurement giving him the outcome $1$ or $0$.

We take the Hilbert spaces under
Alice's and Bob's control initially to be $H_{\rm A}$ and $H_{\rm B}$
respectively, 
and the Hilbert spaces under Alice's and Bob's control at
the end of the protocol to be $H_{\rm A_f}$ and $H_{\rm B_f}$.
The initial and final Hilbert spaces for each party are 
not necessarily identical, since the protocol may require
states to be sent from one party to another. 
By introducing ancillae as necessary, we may take the final
measurements by Alice and Bob to be projective measurements $(P_{\rm A_f} , I_{\rm A_f} - P_{\rm A_f} )$
and  $(P_{\rm B_f} , I_{\rm B_f} - P_{\rm B_f} )$.  

If both parties are honest, the probabilities of all outcomes
are the same in $QP$ as in $P$.    
If Bob is honest, then it cannot be disadvantageous to Alice
to replace $P$ by $QP$: all strategies available to her in 
the former can be replicated in the latter, and the latter
also generally offers her further strategies.   
In particular, the strategy we define below for Alice, when she
does not know $\eta$, has the same success probabilities in $QP$ and
$P$.   We may therefore, for simplicity, without loss of generality assume 
a fully quantum protocol.   
We will write $\eta_{\rm B}$ for Bob's unknown state when 
we wish to emphasize that it is initially under Bob's control. Let 
$S_{\rm AB}$ be the state of all the ancillae introduced by
Alice and Bob in $QP$, and let $U^{\eta}$ be the unitary operation
defined by the protocol up to the final outcome measurements, 
when Alice honestly follows the protocol and believes the state is
$\eta$.   
Here $U^{\eta}$ includes any state transfers between
the parties, as well as local unitaries applied by each party.
Thus $U^{\eta}$ maps $H_{\rm A} \otimes H_{\rm B}$ to $H_{\rm A_f}
\otimes H_{\rm B_f}$.\footnote{We assume the protocol does not require
any states to be discarded, since neither party can trust that
the other will in fact discard states as required. 
Thus $H_{\rm A} \otimes H_{\rm B}$ and $H_{\rm A_f}
\otimes H_{\rm B_f}$ are isomorphic, although the factors
may be different.} 
We have 
\[  \int   \Tr(  ((I_{\rm A_f} -P_{\rm A_f}) \otimes P_{\rm B_f} ) U^{\eta} ( \eta_B \otimes
S_{\rm AB} ) ( U^{\eta})^{\dagger} )  d \mu(\eta) = 1 - \epsilon_{\rm
  C} \, \] 
 where $d \mu( \, )$ denotes the uniform measure over quantum states, and the integral is performed over the entire Hilbert space of $Q_{\rm B}$. 

Now suppose Alice does not in fact know $\eta$. Then she may always
adopt the strategy of choosing a random state $\phi$ from the Hilbert
space of $Q_{\rm B}$ and proceeding with the protocol as if she knows that
$Q_{\rm B}$ is in the state $\phi$.  Since aborting cannot increase $p(1)$, an optimum strategy to maximise $p(1)$
is never to abort, i.e. to take $P_{\rm A_f} = 0$. 
Assuming that Bob always performs his
part of the protocol honestly, the expected value of $p(1)$ is then
\[  \int \int  \Tr( (I_{\rm A_f} \otimes P_{\rm B_f} ) U^{\phi} ( \eta_B \otimes
 S_{\rm AB} ) ( U^{\phi} )^{\dagger} )  d \mu(\eta) d \mu(\phi) \leq
 \epsilon_{\rm S} \, . \] 

Moving the integral inside the trace, and noting that $\int \psi d
\mu(\psi) = \frac{1}{d} I_{\rm B} $, 
where $I_{\rm B}$ is the $d$-dimensional identity matrix on
  $Q_{\rm B}$, we obtain   
 \[ \frac{1}{d} \int  \Tr( ( I_{\rm A_f} \otimes P_{\rm B_f} ) U^{\phi} ( I_{\rm B} \otimes
 S_{\rm AB} )   ( U^{\phi} )^{\dagger} )     d \mu(\phi) \, \leq
 \epsilon_{\rm S} . \] 
The left hand side is bounded below by
\[ \frac{1}{d} \int  \Tr(  (I_{\rm A_f}  \otimes P_{\rm B_f}  )
U^{\phi} ( \phi_B \otimes  
 S_{\rm AB} ) ( U^{\phi} )^{\dagger}  ) d \mu(\phi) \, ,  \] 
which is bounded below by 
\[ \frac{1}{d} \int  \Tr( ( (I_{\rm A_f} -P_{\rm A_f}) \otimes P_{\rm B_f}  ) U^{\phi} ( \phi_B \otimes
 S_{\rm AB} ) ( U^{\phi} )^{\dagger}  ) d \mu(\phi) = \frac{1}{d} (1 - \epsilon_{\rm C} ) \, .  \] 
Hence $\frac{\epsilon_{\rm S}}{1 - \epsilon_{\rm C}} \geq \frac{1}{d}$
as required. 

\end{proof}

\section{Classical protocols}

As noted in the main text, Horodecki et al.\@ \cite{HorodeckiZKP} argue, inter alia, that
non-trivial zero-knowledge classical A-to-B 
protocols with $\epsilon_{\rm C} = 0$ are impossible for a qubit.
For such protocols, Alice must predict some measurement outcome,
and any measurement prediction that holds with certainty for 
a pure qubit $\eta$ and is not certain for a random qubit 
allows Bob to identify $\eta$ exactly,
and so has $\epsilon_{\rm K} = 1$. 
Horodecki et al.\@ did not analyse protocols involving
qudits and also did not consider protocols with
$\epsilon_{\rm C} > 0$.   One way of realising such protocols is for
Alice to choose a projective 
measurement including a randomly chosen projector $P$ from those
with $\braopket{\eta}{P}{\eta} \geq 1 - \epsilon_{\rm C}$ and 
predict the likeliest outcome to Bob. 

Clearly, however, any qudit protocol in which Alice simply sends
Bob a classical prediction of a measurement outcome has 
similar issues to the qubit protocols which Horodecki et al.\@ analysed\cite{HorodeckiZKP}.
Whenever the prediction is highly likely if $Q_{\rm B}$ is in
state $\eta$ and not so likely if $Q_{\rm B}$ is in
a random state, Bob can obtain a significant amount of
information about $\eta$ simply by examining the prediction
and calculating the set of states for which it is highly likely. 
Moreoever, if Bob assumes that Alice is honest, he may choose to refrain from performing
the measurement necessary to confirm her prediction and instead
carry out some other measurement, thus gaining 
additional information.

Horodecki et al.\@ also did not consider relativistic protocols.
Moving to the setting of relativistic quantum cryptography
provides ways to strengthen prediction protocols somewhat, 
since Alice can use secure relativistic bit commitments
\cite{kentrel,kentrelfinite, bcsummoning, AdlamKent1, AdlamKent2}
to commit her predictions, conferring two potential
advantages.
First, Alice can commit to more
than one outcome and subsequently reveal to Bob only the outcome
corresponding to the result that he has obtained, thus decreasing the
amount of information that Bob obtains simply from the fact that she
has made a certain prediction.
Second, Alice need not reveal her
prediction to Bob unless he first tells her the predicted outcome.
The intuition is that this essentially forces Bob to carry out 
(something close to) the specified measurement if he wishes to
have a significant chance of getting information from Alice.
This prevents him from carrying out
a different measurement that gives him (much) additional information.

We now consider classical protocols which use relativistic bit
commitment protocols for one or both of these purposes. 

\vspace{5mm}		
{\bf Classical relativistic KCEKQS protocol 1:} 

There is some agreed frame in which the agents are 
agreed to remain at approximately the same position coordinates
throughout the protocol, respecting the configurations given below. 
We take space and time coordinates with respect to this frame.
We work in units with $c=1$ throughout this paper. 
			
\begin{enumerate}
	
\item Alice and Bob each have two agents, $A_1, A_2$ and $B_1 $,
  $B_2$, configured so that $d = d(A_1, B_1 ) \approx d(A_2, B_2 ) \ll
D=  d(A_1, B_2 ) \approx d(A_2, B_1)$, as in the relativistic bit
commitment protocol of Ref. \cite{kentrelfinite}. 			
			 
\item Alice chooses a projective measurement $\{ P_i \}$ such that
  $\exists x : \Tr(P_x \eta ) \geq 1 - \epsilon_{\rm C}$.
The value of $x$ is secretly shared by both her agents.   
				
\item  
At $t=0$ Alice's agent $A_1$ tells Bob's agent $B_1$ the measurement $\{ P_i \}$.
Also at $t=0$, Alice's agent $A_2$ and Bob's agent $B_2$ initiate a
relativistic bit string commitment protocol committing $A_2$ to the binary
encoding of the index $x$.
				
\item At $t= \delta \ll D$,
$B_1$ performs the measurement $\{ P_i \}$ on $Q_{\rm B}$ and reports
his result to $A_1$.
				
\item At $t = \delta'$, where $\delta' > \delta $ and $\delta' \ll D$, if
 $B_1$ reported the result $P_x$, $A_1$ unveils the commitment
made by $A_2$.   If not, the commitment is not unveiled.   
				
\item If $A_1$ unveils a commitment that matches $P_x$, Bob accepts
the proof, after verifying the unveiled commitment by collecting
data from his two agents.    Otherwise, he rejects.  
				
\end{enumerate}

{\bf Note:} \quad The time coordinates given for this
and later relativistic protocols are merely examples of 
possible timings.  The key requirement is  
that the timings should ensure that at each step, 
relativistic signalling constraints ensure that the 
relevant agent of Alice can have no information about 
data supplied to the other agent by Bob's corresponding
agent at the previous commitment round (if any).      
	
{\bf Security discussion:} \quad 		
First, note that there is no exchange of quantum information
in this protocol.   For the purposes of security analysis, it can
be treated as a relativistic bit string commitment protocol, in
which Alice commits to the bit string defining $x$ via
parallel repetition of the protocol of Ref. \cite{kentrelfinite}. 
For Alice
to cheat, she must be able to unveil the possible bit strings $y$ with 
success probabilities $p_y$ for which $ \sum_y p_y $ is significantly
greater than $1$.   
Conversely, to prove it secure against Alice requires showing that
the relativistic bit string commitment 
protocol ensures that $\sum_y p_y \leq 1 + \epsilon(N, d)$,
where $N$ is the protocol security parameter and
$\epsilon(N,d) \rightarrow 0$ as $N \rightarrow \infty$
for each finite $d$.   

A full proof of security against Alice for this protocol
thus requires analysis of the behaviour of the bit commitment
protocol of ref \cite{kentrelfinite} under parallel repetition. 
The security proof
of \cite{kentrelfinite} for 
of single round classical relativistic bit commitment holds
for both classical and quantum attacks.   
The extension of this argument to bit strings is an interesting topic
in its own right, and we leave discussion of this for future work.

In the present discussion we simply assume without proof that, in the limit as the
security parameters for the bit commitments become large, Alice can
effectively commit to only one outcome of the measurement (or to some
convex combination of outcomes if she makes a probabilistic
commitment).   
Under this assumption, if Alice has
no information about the state $\eta$, her optimal strategy is simply
to commit to a randomly chosen outcome.
Hence the value of $\epsilon_{\rm S}$ is $\frac{1}{d}$, and so
$\epsilon_{\rm S} \rightarrow 0$ as $d \rightarrow \infty$.

{\bf Comment:} \quad 
On the other hand, the protocol does not provide a 
good trade-off between completeness and knowledge concealment. Assuming
Alice does indeed know the state $\eta$, the protocol
succeeds with probability at least $( 1 -
\epsilon_{\rm C} )$.  In this case Bob learns the
value of a projection $P_x$ with $\Tr(P_x \eta ) \geq 1 - \epsilon_{\rm
C}$.

Hence 
\begin{equation}  \epsilon_{\rm K}  \geq  (  1 - \epsilon_{\rm C}
  )^{2} \, .  \label{label1} \end{equation}  
This is a very poor trade-off; in particular 
$\epsilon_{\rm C} \approx 0$ implies $\epsilon_{\rm K} \approx 1$. 

For comparison, if Bob does not take part in the protocol but
instead carries out operations on $Q_{\rm B}$, the maximum expected
fidelity of his estimate for $\eta$ is $\epsilon_{\rm M} =
\frac{2}{d+1}$.  For $\epsilon_{\rm C} \approx 0$ we thus
have $\epsilon_{\rm K} - \epsilon_{\rm M} \gtrsim \frac{d-1}{d+1} \geq \frac{1}{3}$,
and for $d$ large we have $\epsilon_{\rm K} - \epsilon_{\rm M} \approx
1$. 

\vspace{5mm}
One way to improve on the bound (\ref{label1}), at the price of 
increasing $\epsilon_{\rm S}$, is to 
allow Alice to commit to more than one outcome of the projective
measurement.  We now consider a protocol of this type.  

\vspace{5mm}		
{\bf Classical relativistic KCEKQS protocol 2:} 

As with the previous protocol, 
there is some agreed frame in which the agents are 
agreed to remain at approximately the same position coordinates
throughout the protocol, respecting the configurations given below. 
We take space and time coordinates with respect to this frame. 

\begin{enumerate}

\item Alice and Bob each have two agents, $A_1, A_2$ and $B_1 $,
  $B_2$, configured so that $d = d(A_1, B_1 ) \approx d(A_2, B_2 ) \ll
D=  d(A_1, B_2 ) \approx d(A_2, B_1)$, as in the relativistic bit
commitment protocol of Ref. \cite{kentrelfinite}. 		
				
\item Alice chooses a projective measurement $\{ P_i \}$ for which
  there exists a set $S$ consisting of $q$ measurement elements such
  that $\Tr( (\sum_{i \in S} P_i) | \eta \rangle \langle \eta | ) = 
  1 - \epsilon_{\rm C}$ for some agreed value of $\epsilon_{\rm C}$.
The indices $x \in S$ are secretly shared by both
her agents. 

\item 
At $t=0$, Alice's agent $A_1$ tells Bob's agent $B_1$ the measurement $\{ P_i \}$.
Also at $t=0$, Alice's agent $A_2$ and Bob's agent $B_2$ initiate a
relativistic bit string commitment protocols
committing $A_2$ to the binary encoding of each 
index $x \in S$.

\item At $t= \delta \ll D$,
$B_1$ performs the measurement $\{ P_i \}$ on $Q_{\rm B}$ and reports
his result to $A_1$.

\item At $t = \delta'$, where $\delta' > \delta $ and $\delta' \ll D$, if
 $B_1$ reported a result $P_x$ with $x \in S$, $A_1$ unveils the commitment
to $x$ made by $A_2$.   The remaining commitments are not unveiled.  

\item If $A_1$ unveils a commitment that matches $P_x$, Bob accepts
  the proof, after verifying the unveiled commitment by collecting
data from his two agents.    Otherwise, he rejects.   
				
\end{enumerate}

We now show that, conditional on our previously stated
assumptions about the security of relativistic bit commitment
protocols:
			
\[ \epsilon_{\rm K} \geq
q ( \frac{ 1 - \epsilon_{\rm C} }{q} )^2  \qquad {\rm and} \, \,
    \epsilon_{\rm S} = \frac{q}{d} \, . \]

\begin{proof}  
As in the previous proof, we assume that the bit commitment
protocol remains secure under parallel 
repetition.   Hence we limit our discussion to
strategies where Alice commits honestly to a single set of $q$ out of
$d$ possible measurement outcomes in the bit commitment protocol. 

If Alice has no information about the state $\eta$, then she has 
no better strategy than choosing $q$ random orthogonal one-dimensional
projections.   Hence $\epsilon_{\rm S} = \frac{q}{d}$, i.e. $q = \epsilon_{\rm S}  d$.
					
The relativistic bit commitment protocol \cite{kentrelfinite} is
perfectly secure against Bob both for a single bit commitment
and when composed to define a bit string commitment: in either case,
he obtains no information
about Alice's commitments unless and until she unveils them. 
					
Suppose then that Alice knows $\eta$ and randomly chooses a projective measurement $\{ P_i \}$
for which there exists a set $S$ consisting of $q$ measurement
elements such that $\Tr( (\sum_{i \in S} P_i)  \eta  ) = 1 - \epsilon_{\rm C}$.  
If Bob follows the protocol and carries out this measurement, he
obtains outcome $P_i$ with probability $p_i = \Tr (P_i \eta )$. 
If Alice's unveiled commitment confirms that $i \in S$, Bob's
maximum expected squared fidelity guess is $\eta = P_i$, which has 
squared fidelity $p_i $.  
Hence 
\[
\epsilon_{\rm K} \geq \sum_{i \in S} (p_i )^2 \geq
q ( \frac{ 1 - \epsilon_{\rm C} }{q} )^2 = ( 1 -
\epsilon_{\rm C} )^2  q^{- 1} 
\, .
\] 								
\end{proof}
For comparison, if Bob does not take part in the protocol but
instead carries out operations on $Q_{\rm B}$, the maximum expected
fidelity of his estimate for $\eta$ is $\epsilon_{\rm M} =
\frac{2}{d+1}$.  For $\epsilon_{\rm C} \approx 0$ and $q \ll
\frac{d}{2}$ 
we thus have $\epsilon_{\rm K} \gg \epsilon_{\rm M}$. 

{\bf Comment} \quad  A variation on this protocol is for
Alice to randomly choose a projective measurement $\{ P_i \}$
from among those for which there exists a set $S$ consisting of $q$ measurement
elements such that $\Tr( (\sum_{i \in S} P_i)  \eta  ) \geq 1 -
\epsilon$.  She could, for example, use the uniform measure on 
the complete projective decompositions satisfying this criterion.
This variation has $\epsilon_{\rm C} < \epsilon$, 
since Alice's average success probability, when she knows 
$\eta$, is greater than $1 - \epsilon$.   
Similarly, it has $\epsilon_{\rm K} > ( 1 - \epsilon )^{2}
q^{- 1} > ( 1 - \epsilon_{\rm C} )^{2}  q^{-1} $.

             \section{Quantum Alice-to-Bob protocols }

             \emph{Throughout this section, we use the notation $w(a,
b) = \binom{a + b - 1}{a} $.}

             \vspace{2mm} 

             We first show that  the value of $\epsilon_{\rm S}$ for
the quantum Alice-to-Bob protocol set out in the main article is
$\frac{1}{N + 1} + \frac{N}{d(N + 1)}$. 

                    If  Alice does not in fact know $\eta$, the
probability that Bob's measurement obtains a positive result is
maximized if she chooses a single state $\phi$  and prepares all $N$
systems $\{ S_i \}$ in the state $\phi$.  

                    \vspace{2mm}  We use the fact that $\Tr( \Pi_S
(|\phi \rangle \langle \phi|^{\otimes N}\otimes \mathbb{I}^{\otimes M}
) \Pi_S) =  \frac{w(M + N, d)}{ w(N, d)}$\cite{Scaranicloning}. Thus
when  Alice employs the optimum strategy, the probability that Bob 's
measurement obtains a positive outcome is given by:

                    \begin{multline} \epsilon_{\rm S} = \int \Tr( \Pi
(|\phi \rangle \langle \phi|^{\otimes N} \otimes \psi ) \Pi) d
\mu(\psi) = \Tr( \Pi ( |\phi \rangle \langle \phi|^{\otimes N} \otimes
\frac{\mathbb{I}}{d}) \Pi) = \frac{w(N + 1,d)}{w(N,d) d} \\ =
\frac{1}{N + 1} + \frac{N}{d(N + 1)}
\\ \label{lowerbound} \end{multline}

In particular, for $N=1$, we have 		 		
		 		
		 		\[ \epsilon_{\rm S} =  \frac{1}{2} + \frac{1}{2 d} \] 
and $\epsilon_{\rm S} > \frac{1}{2}$ for any $d$.

We can also bound $\epsilon_{\rm K}$.  If Alice follows the
protocol honestly, Bob has available $N+1$ copies of the state $\eta$.
His average squared fidelity between the true state $\eta$ and the best possible
guess from measuring $N+1$ states is
$\frac{N +2}{N + 1+ d}$\cite{Bruss}. Thus $\epsilon_{\rm K} = \frac{N
  + 2}{N+1 + d}$.   Similarly, setting $N=0$, we have that Bob's
optimal
average squared fidelity guess from the original copy of $\eta$ is $\epsilon_{\rm
  M} = \frac{2}{d+1}$.   

From equation (\ref{lowerbound}), we obtain $\epsilon_{\rm K} >
\frac{1}{d \epsilon_{\rm S}}$
and  $\epsilon_{\rm K} - \epsilon_{\rm M} >
\frac{d-1}{d+1} \frac{N}{N+2} \frac{1}{d \epsilon_{\rm S}}$.    
In particular, near 
CS-optimality ($ \epsilon_{\rm S} \approx \frac{1}{d}$) requires 
$N$ large, which implies near-zero concealment ($\epsilon_{\rm K} \approx 1$)
and significant knowledge gain ($\epsilon_{\rm K} - \epsilon_{\rm M}
\approx \frac{d-1}{d+1}$).   For large $d$, this also implies near-complete
knowledge gain ($\epsilon_{\rm K} - \epsilon_{\rm M}
\approx 1$).  

Note that Bob can both follow this protocol honestly in order
to gain evidence about Alice's knowledge of the quantum state, and
then also afterwards attempt to estimate $\eta$ from the $N+1$
states in his possession since the states of these copies of
$\eta$ are not changed in the course of a protocol in which both
parties are honest.
Thus, if Alice honestly follows the protocol, Bob can obtain evidence
of her knowledge of $\eta$ and still 
attain the value of $\epsilon_{\rm K} = \frac{N +2}{N + 1+ d}$
above. 
	 
\section{Quantum Bob-to-Alice protocol}

We begin with a complete description of the quantum B-to-A protocol
summarised in the main text, including the configuration of agents
required for the relativistic bit commitment sub-protocols.   
As with the earlier protocols, there is some agreed frame in which the agents are 
agreed to remain at approximately the same position coordinates
throughout the protocol, respecting the configurations given below. 
We take space and time coordinates with respect to this frame. 
 
\begin{enumerate}

\item Alice and Bob each have two agents, $A_1, A_2$ and $B_1 $,
  $B_2$, configured so that $d = d(A_1, B_1 ) \approx d(A_2, B_2 ) \ll
D=  d(A_1, B_2 ) \approx d(A_2, B_1)$, as in the relativistic bit
commitment protocol of Ref. \cite{kentrelfinite}.

\item Alice and Bob agree in advance on positive integer
security parameters $N$ and $q$. 
	
\item Bob gives $Q_{\rm B}$ to $B_1$, who prepares $N $ additional 
quantum systems $\{ S_i \}$ in states chosen
uniformly at random.
	
\item $B_1$ randomly permutes the systems $\{ S_i \}$ and the system
$Q_{\rm B}$, assigns them all indices from $1$ to $N + 1$, and then gives
all $N + 1$ systems, labelled by their indices, to $A_1$.
	
\item $A_1$ carries out the projective measurement
    $\{ \eta  \, , \,  \mathbb{I} -  \eta \}$ on each of the $N + 1$ systems that
  $B_1$ gave her.   Write $C'$ for the list of indices for
which she obtains outcome $ \eta$; 
let $| C' | = q'$.  If $q' \leq q$, she forms a list $C = C' \cup D$, 
where $D$ is a list of $(q-q')$ copies of the dummy index $0$.\footnote{This dummy index
prevents cheating strategies in which Bob uses the number of Alice's
commitments, made at the next step, to extract additional 
information about the state.}.  If $q' > q$, she picks a random size $q$
sublist $C$ of $C'$.   
	
\item $A_1$ randomly permutes $C$.   
At $t=0$, $B_1$ and $A_1$ initiate
$q$ relativistic bit string commitments \cite{kentrelfinite} committing $A_1$ 
to each of the indices in the permuted list.  Each bit string commitment is set up
so that Alice can commit to any index in $\{ 1, 2,  \ldots N+1\}$.   
This commitment is sustained by $B_2$ and $A_2$ at time $t=\delta$,
where $0 < \delta \ll D$.   These commitments involve a further
security parameter $N'$. 
	
\item At $t= \delta'$, where $0 < \delta' \ll D$ (and for definiteness
we may take $\delta' > \delta$)  $B_1$  tells $A_1$ the index $x \in \{ 1, \ldots, N+1 \}$ that he assigned to $Q_{\rm B}$.
	
\item If $x \in C$, $A_1$ unveils her commitment to that index
(which she initiated and $A_2$ sustained).  
Otherwise she announces failure (or aborts, if the protocol 
includes an abort option) and Bob rejects.\footnote{In the
ideal error-free case, if Alice knows $\eta$ precisely and both
parties honestly perform the protocol, failure is 
possible if and only if $q'>q$.} 
	
\item If Alice's unveiled commitment is indeed $x$, Bob accepts,
once he is able to verify this by comparing data from his agents.   
Otherwise he rejects.

\end{enumerate}
		
\subsection{Security against Alice}
	
Alice's most general possible strategy starts by pre-sharing some 
quantum state between $A_1$ and $A_2$.  $A_1$ receives additional
quantum systems from $B_1$, namely $(N+1)$ qudits. 
She can then carry out any quantum operation on the quantum systems
in her control, depending on the classical data sent by $B_1$,
to generate responses to $B_1$ that purport to initiate the
bit commitment protocols.   $A_2$ can similarly carry out any quantum
operation on the quantum systems
in her control, depending on the classical data sent by $B_2$,
to generate responses to $B_2$ that purport to sustain the protocols.
$A_1$ can then carry out any further quantum operation on the quantum
systems under her control, depending on the index $x$ sent by $B_1$, 
to generate data that purport to unveil $x$ as one of her committed
strings. 

A full security analysis against Alice thus requires a discussion 
of security for a protocol composed of two rounds of a series of
relativistic bit commitment sub-protocols together with the remaining steps
of the protocol above.   We leave this discussion for a future
more general analysis of protocols within which
relativistic bit commitments are suitably composable.
For the present discussion, we will make the restrictive 
assumption that Alice honestly follows each relativistic bit
commitment protocol \cite{kentrelfinite} for the first two 
rounds and uses no quantum information in these relativistic
bit commitment protocols.   Thus we assume that $A_1$ and $A_2$ have pre-shared
classical random numbers which they use, following the protocol
\cite{kentrelfinite}, to commit Alice to some definite classical
bit value in the first round of each bit commitment, and 
to sustain each of these commitments in the second round.   
 
Given that $A_1$ and $A_2$ follow this honest strategy to commit 
and sustain the commitment to a definite classical bit value $b$
for the first two rounds, let $p_S (\bar{b})$ be the probability (for
any unveiling strategy $S$) that
$A_1$ can successfully unveil the opposite
bit value $\bar{b}$ at the unveiling (third) 
stage.   
Each individual relativistic bit commitment used in the
protocol has the property \cite{kentrelfinite} that $p_S (\bar{b})  \leq 
\epsilon(N')$, where $\epsilon(N') \rightarrow 0$ as the security parameter
$N' \rightarrow \infty$.   
We will assume $N'$ is very large compared to $N$ and neglect terms of 
order $\epsilon(N')$ in the following discussion.  (Note that the equalities
and inequalities proven are thus correct only up to order
$\epsilon(N')$).   In this limit, and modulo our
assumptions, the individual relativistic bit commitments are perfectly secure 
against Alice.   Given her restricted strategy, her only
options are to honestly commit to strings of bits $\{ b_i \}$ and
then either to unveil any given bit commitment, revealing the
actual committed bit value $b_i$, or to decline to unveil it.    
       
\subsection{Soundness}

 We now argue that, given these assumptions,
if the protocol involves an unknown $d$-dimensional
state, and we take $(N+1) = Md$ for integer $M$, and
$q = \frac{N+1}{d} = M$, then $\epsilon_{\rm S} = \frac{1}{d}$. 

\begin{proof} 

Alice can achieve a success
probability of $\frac{q}{N + 1}$ by simply committing to $q$
random distinct indices, and hence $\epsilon_{\rm S} \geq \frac{q}{N + 1}$.  We now
show that also $\epsilon_{\rm S} \leq \frac{q}{N + 1}$, and hence
$\epsilon_{\rm S} =\frac{q}{N + 1}$. 

Suppose that $A_1$ and $A_2$ begin the KCEKQS protocol with no
information whatsoever about $\eta$, and that Bob honestly follows
the protocol.   From Alice's perspective, she simply receives 
$(N+1)$ random pure qudits, since $\eta$ is a random qudit and 
the remaining qudits are independently randomly chosen by Bob.
Thus until $B_1$ gives $A_1$
classical information about the index assigned to $Q_{\rm B}$, $A_1$ and
$A_2$'s state of knowledge is exactly symmetrical with respect to all of the $N + 1$
qudits sent by $B_1$. 
$A_1$ is required to initiate commitments, and $A_2$ to sustain them,
before they receive the index assigned to $Q_{\rm B}$, and therefore their
commitment strategy on these rounds cannot depend on that index. 

Thus we may without loss of generality calculate
Alice's success probability by considering some fixed commitment
strategy and averaging over all $N + 1$ possible values for the index
of $Q_{\rm B}$. For simplicity, we will analyse a related protocol $T$ in
which Alice always makes $q$ commitments and subsequently unveils all
$q$ commitments.   We will say that Alice is 
successful in $T$ if and only if at least she unveils at least one
valid commitment to the correct index for $Q_{\rm B}$ (whether or not other
unveiled commitments turn out to be valid commitments to any index).
Alice's success probability $\epsilon_{\rm S}$ in the KCEKQS protocol is no greater than her success
probability in $T$, since in either case Alice succeeds if and only if
she can unveil at least one commitment to the correct index.

Now, for the purpose of obtaining a
contradiction, suppose that Alice can succeed in protocol $T$ with
probability greater than $\frac{q}{N + 1}$. If the probability
distribution over the values of Alice's unveiled commitments in task $T$
depends only on her initial commitment strategy and not on the
classical information she subsequently obtains from Bob, then the values
of her $q$ unveiled commitments must be uncorrelated with the index of
$Q_{\rm B}$.  Therefore the probability that one of these $q$ unveiled commitments is to
the correct index for $Q_{\rm B}$ can be no greater than $\frac{q}{N + 1}$,
with this bound saturated whenever Alice uses a strategy which always
produces $q$ valid commitments to different bit values in $\{1, 2,
... N + 1\}$.  Thus if the success probability is greater than
$\frac{q}{N + 1}$, the probability distribution over the values of her
unveiled commitments must be correlated with the classical information that she
receives from Bob.   

If so, under our assumptions about
Alice's restricted strategy, Alice must have some freedom
to choose whether to unveil $0$ or $1$ for at least one bit $i$ out of the
$\approx q \log (N+1) $ bits to which she commits.
More precisely, for at least two
different indices $q, r$ in $\{1, 2, \ldots ,  N + 1\}$, if $p_i( 0 | q)$ is
the probability that Alice unveils a $0$ for bit value $i$ when Bob
tells her the index of $Q_{\rm B}$ is $q$, and $p_i(1 | r)$ is similarly defined,
then $p_i(0 | q) + p_i(1 | r) > 1$.   But this contradicts the security of the relativistic bit
commitment protocol \cite{kentrelfinite}, under our assumptions about
Alice's restricted strategy.   (Recall again that we neglect terms of 
order $\epsilon(N')$ in this discussion.) 
Hence, given those assumptions, Alice
cannot succeed in protocol $T$ with probability greater than
$\frac{q}{N + 1}$.   Hence, again given those assumptions, in our KCEKQS protocol, $\epsilon_{\rm S} \leq \frac{q}{N +
1}$. 
\end{proof}
	
\subsection{Completeness}
We now show that $\epsilon_{\rm C} \rightarrow 0$  as $N \rightarrow \infty$.
\begin{proof}
If Alice does know the state $\eta$, she will 
get a positive outcome on $Q_{\rm B}$. 
The probability
that her commitment fails to be accepted is thus given by:
\begin{equation}\label{epsc}
\epsilon_{\rm C} = \sum_{x = q }^{N} P( X_{N} = x) \frac{x + 1 - q}{x + 1} \, ,
\end{equation}
where $p(X_{N} =x)$ is the probability that $x$ out of Alice's ${N}$
measurements on the $N$ systems $S_i$ obtain the result $\eta $.
		
If Bob is honest and chooses the states of the systems $S_i$ at
random, the result $ \eta $ is obtained on each
run with probability $\frac{1}{d}$.
Hence $X_N$ is binomially
distributed:
\begin{equation}
P(X_N = x) = \binom{{N}}{x} \frac{1}{d^x}(1 -
\frac{1}{d})^{{N} - x} \, .
\end{equation}
For $q = \frac{N + 1}{d}$, the distribution
$X_N$ has mean $q  - \frac{1}{d} = \frac{N}{d} $.   
Hoeffding's inequality implies that 
\begin{equation} \label{hoeffding}
P (X_N \geq \frac{N}{d} + \epsilon N ) \leq \exp ( - 2 \epsilon^2 N )
\, .
\end{equation}
It follows from equation (\ref{epsc}) that
\begin{equation}
\epsilon_{\rm C} \leq \epsilon d p_{\rm mode} + \exp ( - 2 \epsilon^2 N )
\end{equation}
for any $\epsilon > 0$, where $p_{\rm mode}$ is the maximum over $x$
of the binomial distribution $P (X_N = x )$.
We can obtain an adequate bound simply by using $p_{\rm mode} \leq 1$.\footnote{
A tighter bound for $N \gg d$ follows from the normal approximation
to the binomial distribution, which gives 
$
p_{\rm mode} \approx \frac{1}{\sqrt{2 \pi}} \frac{1}{\sqrt{N
    \frac{1}{d} ( 1 - \frac{1}{d} ) }}$ . }

Taking, for example, $\epsilon = \frac{1}{2} N^{-\frac{1}{2}} ( \log N
)^{\frac{1}{2}}$, we see that 
$\epsilon_{\rm C} \rightarrow 0$ as $N \rightarrow \infty$.  

\end{proof} 

\subsection{Knowledge concealment}

Bob begins with one copy of $\eta$.   If dishonest, he may combine
this with any ancillae he wishes, carry out any quantum operations
he wishes, and produce a (perhaps highly entangled) state including
$(N+1)$ qudits that he sends to Alice, together with a system 
that he retains.  He may then carry out any measurement he wishes
to produce an index $x$.    Alice responds, effectively, with either
a $1$ (if she unveils a commitment to $x$) or a $0$ (if she fails
to unveil a commitment to $x$).   Bob may then carry out any 
measurement he wishes on his retained system to produce his 
maximum possible expected squared fidelity estimate of $\eta$. 
This measurement choice may depend on Alice's response.    

Effectively, Bob's task is to optimize his state estimation 
in this scenario.   His overall strategy $S$ is fixed up to
Alice's response, but then may involve different
state estimation strategies depending on the one bit of information
supplied by Alice.  
For any given overall strategy $S$, the overall expected squared fidelity
of his estimate is 
\begin{equation}
p_S (0) f_S (0) + p_S (1) f_S (1) \, , 
\end{equation}
where $p_S (b)$ is the probability of outcome $b$ given strategy
$S$ and $f_S (b)$ is the expected squared fidelity obtained 
from $S$ conditioned on outcome $b$.  

Suppose that Alice does not respond at all.
Bob may simply follow the fixed strategy $S_b$ given by 
following $S$ and assuming outcome $b$. 
This produces an expected squared fidelity of at least
$p_S (b) f_S (b)$.   Since $S_b$ is a possible strategy
for the standard task of state estimation given one copy
of an unknown qudit (and no further information), 
we have that
\begin{equation}
p_S (b) f_S (b) \leq f_{\rm max} = \frac{2}{d+1} \, 
\end{equation}
where the right hand side is the maximum expected square fidelity
obtainable from any state estimation strategy on an unknown qudit
\cite{Bruss}. 
Hence Bob's overall expected squared fidelity
\begin{equation}
p_S (0) f_S (0) + p_S (1) f_S (1)  \leq \frac{4}{d+1} \, . 
\end{equation}
That is, $\epsilon_{\rm K} \leq \frac{4}{d+1}$.   
For comparison, we have that Bob's optimal expected squared fidelity if 
he does not participate in the protocol but simply carries out
measurements on $\eta$ is $\epsilon_{\rm M} = \frac{2}{d+1}$.  
Thus, for large $d$, the protocol is highly knowledge concealing and
close to zero knowledge.

{\bf Comment} \quad 

Our discussion to date has been based on the assumption that the state
$\eta$ is pure.   This excludes the possibility that
Alice has access to one or more systems which are entangled with
$Q_{\rm B}$.    However, another natural scenario (call it ${\rm Sc}_1$) in which a KCEKQS protocol might
be useful is that $Q_{\rm B}$ is maximally
entangled with a system $Q_{\rm A}$ in Alice's possession, and that Alice
knows the joint state.  

One option for Alice is then to perform a
projective measurement on $Q_{\rm A}$ before the start of the protocol, so
that the system $Q_{\rm B}$ is subsequently in a pure state $\eta$ known
to her.   This recreates our original scenario (call it ${\rm Sc}_0$). 
Thus, if Alice can succeed in any given KCEKQS protocol
with probability $(1- \epsilon_{\rm C})$ if she knows $\eta$, 
she can also succeed with probability at least $(1- \epsilon_{\rm C} )$ 
in the scenario of the previous paragraph.   
Conversely, while there are clearly protocols that give Bob evidence
favouring ${\rm Sc}_1$ over ${\rm Sc}_0$, 
no KCEKQS protocol can give Bob evidence favouring ${\rm Sc}_0$ over ${\rm Sc}_1$.

{\bf Allowing an abort option} \quad 

Our relativistic quantum Bob-to-Alice 
protocol could be altered to achieve $\epsilon_{\rm C} = 0$
by allowing Alice to abort the
protocol whenever she obtains a positive outcome more than $q$ times
at step five \footnote{One could also include an abort option for
the case where  Alice does not obtain a positive outcome for any one
of these measurements.   However, there is a significant distinction.
In that case, in the error-free model, Alice knows for certain that
Bob is trying to cheat.  Since announcing an abort 
leaves open the possibility that Bob was honest and a statistically
unlikely outcome was obtained, she may
prefer to directly state that Bob is cheating, or simply stop
the protocol.}.
An honest Alice with perfect knowledge will abort whenever
the binomial random variable $X_N > q$.  
For $q = \frac{N+1}{d}$ and large $N$, this implies an
abort probability of roughly $\frac{1}{2}$.
Taking $q = \frac{N}{d} + \epsilon N$, with $\epsilon > 0 $, we see from
equation (\ref{hoeffding}) that the abort probability can be 
bounded by $\exp ( - 2 \epsilon^2 N )$.   
This gives a protocol with $\epsilon_{\rm C} = 0$,  $\epsilon_{\rm S} = \frac{1}{d} +
\epsilon$ and with abort probability that tends to zero for large
$N$.    

These parameters may represent a reasonable tradeoff in some 
circumstances.   However, introducing an abort option does not eliminate the
possibility of unjustified mistrust.   
Without an abort option, it is possible that Alice and Bob may both honestly follow
the protocol, and that Alice may know $\eta$ precisely and thus correctly
identify $Q_{\rm B}$ as a candidate, but that Alice may be unable to persuade Bob
of this because she had more than $q$ candidates and her random
choice of a size $q$ subset did not include the index of $Q_{\rm B}$. 
Introducing an abort option removes the possibility of honest Alice being falsely
suspected of cheating by honest Bob \emph{for this specific reason}.
However, an honest Bob may now unfairly suspect an honest Alice of
cheating if she honestly aborts.   A dishonest Alice might abort
because she had no information about $\eta$, carried out no measurements, made random or invalid
commitments, and thus effectively used the protocol 
to steal Bob's copy of $\eta$.   Bob cannot tell whether or not the protocol
has been honestly aborted.  

No version of this protocol can
guarantee that honest Bob accepts the honesty of honest Alice.
This is because KCEKQS is a one-shot procedure, and 
this protocol has nonzero probability of either failure or abort (if
there is an abort option).    Bob has only one copy of
$\eta$, and so if he gives that copy to Alice and the protocol fails
or is aborted, he has no further opportunity to learn anything about
$\eta$ or about Alice's knowledge of it.    
Of course, the parties might be able to repeat the protocol using a new
state, but even if this next protocol succeeds,  Alice will have
proved only that she knows the new state, which gives Bob no 
direct evidence about whether she knew the previous state. 

In summary, in most scenarios, it seems to us that the abort option
version of the protocol gives no clear advantage. 

\section{Further Security Issues}

Here we discuss several potential security issues and possible
weaknesses for KCEKQS protocols which are not covered by our earlier security definitions.

\subsection{What precisely does Alice give evidence of?} 

We have defined security for KCEKQS in terms of
two extremes, the value of $\epsilon_{\rm S}$ giving the probability
that  Alice's proof is accepted if she knows nothing at all about the
state, and the value of $1 - \epsilon_{\rm C}$  giving the probability
that  Alice's proof is accepted if she knows the state exactly. 
As noted above, there are intermediate possibilities.   
For example, Alice could have a significant amount of classical
knowledge about the quantum state without
knowing it exactly; she could have quantum information correlated
with the state (such as a number of copies), without any additional
classical information; she could have beliefs about the 
state encoded in a probability distribution.  Such beliefs 
need not necessarily be well founded or correct: for instance she might
believe the classical description of the state is $\eta' \neq \eta$. 

In each case, if Alice's information or beliefs allow her
to produce a guess $\eta'$ at the classical description of
$\eta$ such that $ \Tr ( \eta \eta' ) \approx 1$, 
she can produce what we have called
``evidence of knowledge''.  
In each of the KCEKQS protocols described
above, Bob will accept this ``evidence of knowledge'' with probability close to $1 -
\epsilon_{\rm C}$ -- even though Alice may not only
not have complete knowledge of $\eta$ but may even have
false beliefs about its classical value.   
As this illustrates, the notion of knowledge of a quantum state needs
careful analysis.   One might be inclined to frame a definition
so that if Alice believes the state is the pure state $\eta'$, and it is actually
$\eta \neq \eta'$, then she has no knowledge about the state,
even if $ \Tr ( \eta \eta' ) \approx 1$. 
But if one does, one has to accept that someone who has no
knowledge about a state can nonetheless appear to give strong
evidence of knowledge.   Alternatively, one might frame a definition
so that, if Alice has probabilistic beliefs about the state $\eta$ 
encapsulated in the density matrix $\rho$, then  $\Tr ( \rho \eta )$ 
is a measure of her knowledge.   If so, one has to accept that 
false (not merely uncertain) beliefs are consistent with a high
degree of knowledge.    

Without giving a precise general definition of knowledge of a quantum 
state, our security definitions still establish
that the protocols give evidence of knowledge of $\eta$ in an
interesting, if restricted, sense.
Namely, they
strengthen the evidence for the hypothesis that Alice knows the
precise classical value of $\eta$ compared to the hypothesis that she
has no classical or quantum information correlated with $\eta$.    

One might frame additional security definitions that allow
more to be established.  For example, one could require 
{\it strong non-triviality}: if Alice does not have and is not
able to obtain a precise classical
description of $\eta$, then the probability
that her proof is accepted is strictly less than $(1 - \epsilon_{\rm
  C})$.  This requirement holds for our classical relativistic KCEKQS
protocol $1$  
and for our quantum B-to-A protocol but does not hold for our
classical 
relativistic KCEKQS protocol $2$,
in which Alice can ensure acceptance probability $(1-\epsilon_{\rm C})$ 
even if she only knows a subspace of Hilbert space of dimension
$q=\epsilon_{\rm S} d$ in which the state lies. 
Nor does it hold for our quantum A-to-B protocols, in which Alice
can ensure acceptance probability $(1-\epsilon_{\rm C})$ 
even if she only has some way of producing precisely $N$ copies of
$\eta$ (and no more), and no other classical or quantum information correlated with $\eta$. 

\subsection{Composability and Accumulation of Information}

Zero-knowledge proofs are often used as sub-protocols in
the construction of more complex protocols, such as electronic voting
schemes and digital signature schemes. 
One might hope that our knowledge-concealing protocols could be used
as building blocks for quantum protocols of this type.
One security concern here is composability: our security arguments
hold for single-shot instances of the protocol, and collective 
attacks on combinations of several protocols would need to be
addressed separately. 
Among other things, since the amount of
information obtained by Bob in a KCEKQS protocol will never be
\emph{exactly} zero for any finite-dimensional quantum state, 
one would have to consider carefully whether Bob's
information gains could accumulate in such a way as to undermine the
overall security.   We leave further investigation for future work. 

\subsection{Alternative Measures of Bob's information gain}

Density matrices characterise an
agent's knowledge about the outcomes of measurements that may be made
on a system; they do not completely characterise an agent's knowledge
of how that system was prepared.
Our knowledge-concealing criterion thus does not necessarily 
characterise all the information Bob might learn in the course of
any conceivable protocol.   
As a hypothetical illustration for the case of a qubit, if 
Bob were somehow to learn in the
course of a protocol that the original state was definitely either
the state $| + \rangle$ or the state $| - \rangle$, there is a sense in which
he has gained significant information.   However, the density matrix $\rho$
describing his new state of knowledge and the density matrix describing his
knowledge before the protocol are both the maximally mixed state.
Our fidelity-based knowledge-concealment criterion would thus suggest 
that he has gained no information.  

We believe that $\epsilon_{\rm K}$ is nonetheless a good
measure of Bob's information gain for a single shot protocol,
because it quantifies Bob's ability to predict the outcome of measurements 
on a system in the state $\eta$ and/or to prepare a state that
successfully simulates the state $\eta$.
These operationally defined measures of information about
a state are relevant in most cryptographic contexts and 
are the only relevant measures in some.  

That said, there is no unique measure of information about a quantum
state that characterises every property that is relevant in every
possible scenario.   Even assuming that Bob's aim is to produce a
guess $\eta'$ at $\eta$ that optimizes some cost function, and 
assuming also that the cost function depends only and monotonically on 
$F( \eta, \eta' )$, there are still infinitely many cost functions
that may be considered.   As a simple example, Bob might be given a fixed
reward if and only if $F( \eta , \eta' ) > 1 - \epsilon$ for some
small $\epsilon > 0$.   In this case, a sensible parameter 
might be $\epsilon'_{\rm K}$, defined as the probability of this condition 
holding after the protocol if Bob follows an optimal strategy. 
A  sensible comparison would then 
be to $\epsilon'_{\rm M}$, the probability of it holding 
if Bob instead carries out his optimal strategy that involves 
operations only on the unknown state.  

\subsection{What does  Alice learn?} 

We have considered in detail the amount of information
that Bob gains about the state $\eta$ in the course of our protocols,
but we have not thus far considered how much \emph{ Alice} could
potentially learn about $\eta$ if she does not in fact know this
state.  In some applications it might not matter if  Alice gains
information about $\eta$, but if we are using a KCEKQS protocol
precisely because knowledge of the  state $\eta$  has value in some
context (such as quantum money\cite{wiesner} or quantum
voting\cite{evoting}), then it may be important to limit the amount of
information that might potentially be gained by a dishonest Alice.

Both parties may also be concerned about the information available to an
eavesdropper Eve who intercepts protocol communications.  
In principle, eavesdropping can be completely prevented by using secure 
classical and quantum channels.  Secure classical channels can be 
ensured by using one time pads.   In practice, though, secure quantum channels
require distributing and storing perfect entangled states, so 
that quantum eavesdropping presently remains a practical concern. 
We will assume secure classical channels but potentially insecure quantum channels
below.  

The quantum Alice-to-Bob protocols give Alice no additional 
information.   Bob need not tell 
Alice whether or not his measurement obtained a positive result,
so Alice learns nothing whatsoever from the protocol.   
However, these protocols are vulnerable to quantum eavesdropping, 
since Eve could intercept the state copies sent by an honest
Alice, and obtain information by state estimation.

In our classical protocols, assuming that Bob
performs the protocol correctly and honestly,  Alice can gain 
some information about the state since Bob  must
tell Alice the result of a measurement that he has performed on the
system $Q_{\rm B}$. But clearly since  Alice obtains
information about the state only via Bob's classical communication,
she learns no more than Bob  does in the course of such
protocols, if she begins with no knowledge about the state.\footnote{Alice can however learn more than Bob does if she
begins with some information about the state.}

Our quantum Bob-to-Alice protocol is significantly vulnerable to
a dishonest Alice who wishes to obtain rather than prove knowledge.
If Bob performs this
protocol honestly he learns nothing whatsoever about $\eta$
from the protocol.   However, a dishonest Alice can store all 
the quantum states sent to her and make commitments to random
indices, or fail to make valid commitments at all. 
She can then perform a measurement
on $Q_{\rm B}$ after Bob has told her the index of $Q_{\rm B}$.
Effectively, she can thus steal $\eta$ from Bob and  
obtain as much information about it as he could have 
obtained unaided. 
For large $d$, this strategy has a low probability of successfully persuading Bob
that Alice has honestly followed the protocol and knew the state. 
Nonetheless, Alice dishonestly gains some information about $\eta$, 
whereas an honest Bob sacrifices his copy of $\eta$ and gains
no information at all.    
If Eve intercepts the quantum communications but not the classical
communications, she can steal all the states sent by Bob but will not
learn which one is $\eta$.  If $N$ is large, she can thus obtain very
little information about $\eta$. 

For large $d$, the information a dishonest Alice can gain
is small, assuming she begins with no information; for large $N$,
the information Eve can obtain is even smaller.   
Still, in contexts where even
limited information about $\eta$ is highly valuable, these are 
undesirable weaknesses.    

\subsection{What if Bob has additional information?}

We have calculated bounds on the information that Bob gains during the
course of each protocol assuming he starts with
strictly zero knowledge about the state of $Q_{\rm B}$ and only ever has a
single copy of the state of $Q_{\rm B}$.

In other possible scenarios, Bob might also have some limited
classical information about $\eta$, or have additional
correlated quantum information (for example further copies of $\eta$),
or both.   Our bounds do not necessarily apply in
such scenarios. 
For example, suppose that 
Bob has $  Mq$ copies of $\eta$, for some large $M$, 
and that he and Alice perform classical relativistic KCEKQS protocol
$2$. 
Bob can apply the projective measurement specified by Alice 
on every copy in his possession.  Suppose that Alice is honest, 
and suppose also that $\epsilon_{\rm C} \ll (M q)^{-1}$.  
Bob is then likely to obtain positive outcomes for $r$ elements
of the measurement basis, where $r \leq q$. 
This allows him to identify a subspace $V$ of dimension $r$ 
such that $\Tr (P_V \eta) \approx 1$.